# Neurobiological reality simulation through an Artificial Neural Network at criticality

Yiannis Contoyiannis[(1)] and Myron Kampitakis[(2)]

(1) Department of Electric-Electronics Engineering, West Attica University, 250 Thivon and P. Ralli, Aigaleo, Athens GR-12244, Greece (email: yiaconto@uniwa.gr)
(2) Hellenic Electricity Distribution Network Operator SA, Network Major Installations Department, 72 Athinon Ave., N.Faliro GR-18547, Greece (email: m.kampitakis@deddie.gr)

**Abstract :** An artificial neural network (ANN) based on fundamental principles of physics can simulate the operation of neurobiological reality of membrane potential as well as the properly defined order parameter. This ANN operates in conditions of criticality and simulates the behavior of an excitatory biological neuron, especially the relaxation phase where the critical fluctuations of biological neuron appear. These critical fluctuations can not be explained by the Hodgkin-Huxley (H-H) model. A proposal for the origin of these fluctuations is being discussed.

Simulation of real neuronal function is a valuable research effort in recent years. This simulation can be done by solving the systems of differential equations resulting from the work of Hodgkin-Huxley (H-H) [1]. In general, we could say that the action potential of neurons includes the phase of the spike and is completed with the relaxation phase between two successive spikes (see Fig.1).

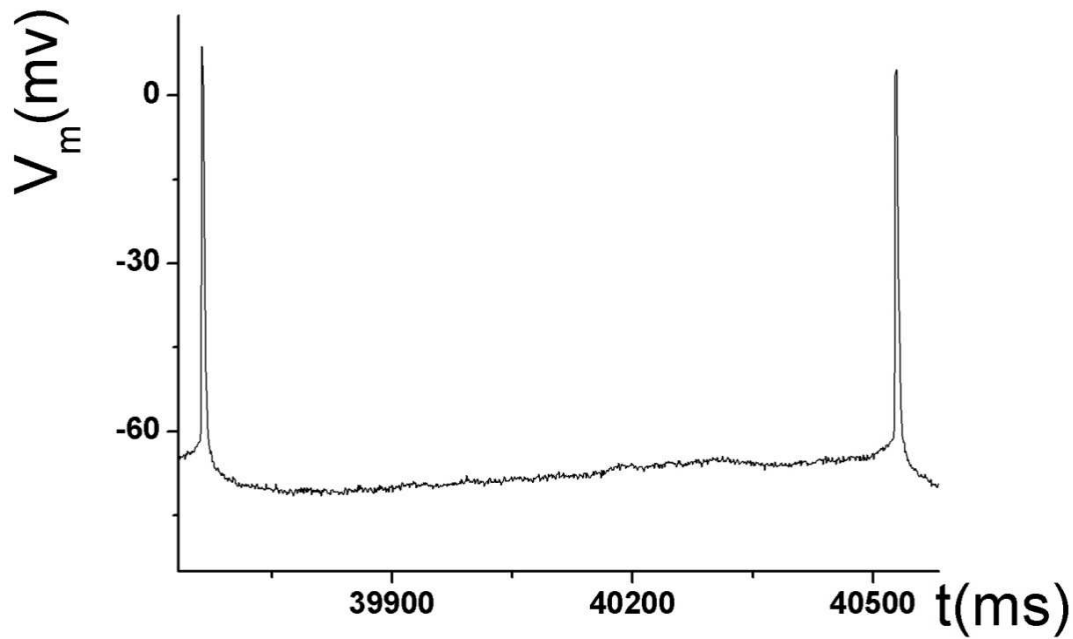

*Fig.1 The membrane potential fluctuations vs time. Two spikes as well as the relaxation phase (grass) are presented.*

While for the spike phase there are no serious doubts today that the H-H model describes what we see very well, for the phase of the relaxation phase the model seems to be unable to describe the corresponding dynamics [2]. Recently [2], we studied this dynamics in biological networks of neurons and concluded for the first time that this dynamics is very well described by the critical dynamics which is a mechanism of critical intermittency. The purpose of the present work is to propose an artificial neural network (ANN) model that successfully simulates the operation of a biological neuron that operates within the environment created by the other neurons connected to it. A description will be successful if it is accomplished to describe both the phase of the spikes and the dynamics responsible for the grass between the spikes. For the phase of spikes, the results are directly comparable to the structure of the spikes of biological neurons, but for the dynamics that creates the grass, the required processing should be done. Today we have a good way to do that. It is the search for the critical situation through the method of critical fluctuations (MCF). Thus, from the artificial neural network, we produce a time series of the relaxation period phase and proceed with the MCF critical analysis in the search for criticality. We compare the results with those of the biological neuron analysis with the MCF that were done in our work [2]. In this work we have shown that the fluctuations of the membrane potential for excitatory neurons follow critical dynamics. Thus, if the simulation gives us similar results, that is critical behavior, and can reproduce the phase of the spikes, then we can indeed claim that our neural network is satisfactorily simulating the neurobiological reality. The model we will present is an artificial stochastic network structured in basic principles of physics, such as the principle of energy minimization as dictated by the second law of thermodynamics. Essentially this is a procedure that could be included in the optimization class [3] in ANN. At the algorithmic level, we follow the Metropolis algorithm by which we study the phase transitions in spins systems (Ising, Potts models) where we have made modifications, basically changing the selection of configurations through Fermi-Dirac Statistics, and not through standard Boltzmann Statistics.

## 2. The proposed ANN

We focus on quantized states in an ANN by considering a network of $n$ neurons, whose output states are random variables $\xi_i, i = 1, \ldots n$ that can take the values +1 or -1 expressing the "excitatory" or the "inhibitory" character of each neuron, respectively. Each neuron of the network is connected with all other neurons comprising an extensive feedback structure spanning over the whole network. Moreover, the connection weights $T_{ij}$ may take either positive or negative values, according to the synaptic properties in the connection between two biological neurons. Then, based to the ANN formalism, the energy function representing the state of the ANN at time $t$ is given as [4].

$$E(t) = -\sum_{i,j=1}^{n} T_{ij}\, \xi_i(t) \xi_j(t) \quad (1)$$

Physical reasons (Metropolis Algorithm) impose us to use quadratic terms like energy function rather than propagation function, $F_i(t) = \sum_{j=1}^{n} T_{ij}\, \xi_j(t)$, in contrast to what happens in a usual ANN. The basic quantity in neural networks is the synapses matrix T. In the elements of matrix, $T_{ij}$ i, j = 1, ... n we give the following properties:

- $T_{ij} = 0$ when i=j ( non self-interaction)
- $T_{ij} \neq T_{ji}$ because this is more close to reality of natural neurons
- According to recent findings from neurophysiology [5] almost the 80% of synapses are excitatory while the rest 20% are inhibatory. Another result is that the intensity of inhibatory synapses is almost 4 times bigger than the excitatory.

The value (+1, -1) for variables $\xi_i, i = 1, \ldots n$ selected by Metropolis Algorithm [6] is determined from the energy value. Therefore, as we see from equation 1, each neuron depends on the global state of the network expressed in terms of the suitably defined total energy in contrast to the spin models where changes of the single spin state are determined by the energy change introduced by the nearest neighbors. This criterion is based on the idea that in multiple connected networks, going beyond the first (or first and second) neighbor interaction, describes the natural situation in a more consistent manner. In our approach, the Metropolis algorithmic time is included in order to implement a basic principle of Physics, the principle of energy minimization as dictated by the second law of thermodynamics. Essentially this is a procedure that could be included in the optimization class [3] of ANN algorithms in the framework of the minimization or maximization of certain objective function(s). Despite this, in the model we propose here, the weight function, used in the Metropolis algorithm, has the form of the Fermi-Dirac distribution for fermionic degrees of freedom. This is in contrast to the spin model case where the Boltzmann weight is exclusively used. In each step of the modified Metropolis algorithm, the flip of a single spin is performed if the energy of the resulting configuration decreases, and with probability determined by the Fermi-Dirac distribution:

$$f = \frac{1}{1 + e^{-\beta \varepsilon}} \qquad (2)$$

which describes the probability of quasi-particle excitations in a Fermi liquid [7]. In Eq. (2) $\beta$ is the inverse of temperature and $\varepsilon$ (which is usually the energy of the single fermion) expresses the energy of the entire system. The probability for the output of neuron $i$ at time $t+1$ to make a transition between the states $\pm 1 \rightarrow \mp 1$ is calculated as:

$$\Pr\left(\xi_i(t+1) = +1\right) = \frac{1}{1 + e^{-\beta E(t)}} \qquad (3a)$$

$$\Pr\left(\xi_i(t+1) = -1\right) = 1 - \Pr\left(\xi_i(t+1) = +1\right) = \frac{1}{1 + e^{\beta E(t)}} \qquad (3b)$$

In this way we make a link between the introduced ANN model and a fermionic system possessing exclusively spin degrees of freedom. Note that in Eqs. (3) we consider that for all neurons the exponent factors $\beta_i = \beta$. This is due to the fact that we assume global equilibrium in the introduced model. Certainly, within the neural network description local equilibrium is a possible scenario. In this case $\beta_i \neq \beta_j$.

We can interpret $1/\beta$ as the temperature of a thermal system [4]. Indeed, as $\beta$ increases, the temperature decreases and as dictated by Eqs. (3) the $\Pr\left(\xi_i(t+1) = +1\right) \rightarrow 1$ while the $\Pr\left(\xi_i(t+1) = -1\right) \rightarrow 0$. This leads to broken symmetry phase for the +1,-1 states. As $\beta$ decreases, the temperature increases, and $\Pr\left(\xi_i(t+1) = +1\right) = \Pr\left(\xi_i(t+1) = -1\right) = \frac{1}{2}$. This leads to a symmetric phase for the +1, -1 states. Note that, the transition between symmetric phase and broken symmetry phase appears in thermal systems as the temperature, which is the control parameter, decreases. Therefore, $1/\beta$ is considered the control parameter for the introduced ANN model. A local field of this transition could be $m_i(t)$, which under the consideration $\beta_i = \beta$ takes the same value for all neurons:

$$m_i(t) = \Pr\left(\xi_i(t+1) = +1\right) - \Pr\left(\xi_i(t+1) = -1\right) = \frac{1}{1+e^{-\beta E(t)}} - \frac{1}{1+e^{\beta E(t)}} = \tanh\left(\frac{\beta}{2}E(t)\right) \quad (4)$$

Finally, it should be clarified that the introduced ANN model has no external inputs. The only "external" influence is the value of the control parameter $1/\beta$.

## 3. The Algorithm steps of the model

From the definition of the $m(t) \equiv m_i(t) = \Pr\left(\xi_i(t+1) = +1\right) - \Pr\left(\xi_i(t+1) = -1\right)$

according those mentioned in the previous section the $m(t)$ quantity has a character of the order parameter as the mean value, therefore magnetization is the order parameter in the thermal critical phenomena. The membrane potential of the biological neuron should correspond to a physical potential of the form $mH_F$ where $m$ is the order parameter and $H_F$ an external field. The role of the extrenal field is played by the T-matrix whose elements are the neuron interactions with the other network neurons that act as an external field. Therefore the matrix elements $[V_{ij}]$ of membrane potential are given as :

$$V_{ij} = T_{ij}\frac{m_j(t)+1}{2} \quad (5)$$

where we replace the order parameter $m_j(t)$ with its average $\frac{m_j(t)+1}{2}$ to avoid its zeroing in the critical state where the order parameter is close to zero. The membrane potential of the $i$ neuron is given as :

$$V_i = \sum_{j=1}^{n} V_{ij} \quad (6)$$

According to the standard formalism of ANN [4] we can deduce an update rule which describes the time evolution of the $i$ neuron in the environment of $j=1,2..n$ neurons :

$$m_i(t+1) = \tanh\left(\frac{\beta}{2}V_i\right) - \varepsilon_i \quad (7)$$

Where $\varepsilon \in [-\varepsilon_o, \varepsilon_o]$ a uniform low amplitude noise necessary to bring the map $\{t\} \to \{t+1\}$ in the conditions of ergodicity [8].

## 4. The simulation

To show that the model can actually reproduce the critical dynamics of the excitatory biological neuron, we have to show that the produced time series of the order parameter of the equation (7) obeys the critical dynamics. The critical state in phase transition indicates the transition from the symmetrical phase to the phase of broken symmetry as discussed previously. In systems consisted of random variables which take the values +1 or -1  like the Ising  spin models, this occurs at a critical temperature. In two dimensional systems as the 2D-Ising model the critical temperature has the value (in arbitrary unit system) $T_c = 2.3$ [9] and the average magnetization is zero. The corresponding inverse temperature is $\beta = \frac{1}{T} = \frac{1}{2.3} = 0.4348$.  To determine the critical state we will consider a quantity produced from $m(t)$  called effective  order parameter and defined as:

$$M = \frac{FS^{(+)} - |FS^{(-)}|}{FS^{(+)} + |FS^{(-)}|} \quad (8)$$

with $FS^{(+)} = \sum_{t=1}^{N_{iter}} m^{(+)}(t)$, $FS^{(-)} = \sum_{t=1}^{N_{iter}} m^{(-)}(t)$ where $m^{(+)}(t)$ and $m^{(-)}(t)$ are the positive and negative values of the $m(t)$ time-series respectively, and $N_{iter}$ the iterations number.  As the effective order parameter M becomes close to zero , the system is closer to criticality. The data in the numerical experiment   are  n = 20 neurons, $N_{iter} = 297000$, $\varepsilon_o = 0.0004$.  At the temperature T = 2.3 we change the initial conditions to achieve a M value close to zero. Here we found an initial condition for which M = 0.017. There are a large number of initial conditions that give M values at the order of $10^{-2}$ which we can also use. We study the dynamics of the neuron $i = 1$ but we can also study any of the twenty neurons. The produced time-series consists from a grass around zero and from spikes emerging from zero grass towards the positive values. This structure refers at 50% of the neurons. The other 50% neurons give time-series consisting of the grass around zero but now the spikes emerge towards negative values. Such a structure is presented for neuron $i = 2$. So the zero appears as a fixed point as we expect for the order parameter. Therefore in Figure 2 we present the two cases as time-series, in which the values are located at a zone close to zero  for e.g [-0.03,+0.03]. The produced time-series are applications of equation (7) for $i = 1,2$ :

$$m_1(t) = \tanh\left(\frac{\beta}{2} V_1\right) = \tanh\left(\frac{\beta}{2} \sum_{j=1}^{n} T_{1j} \frac{m_j(t)+1}{2}\right) - \varepsilon \quad (9)$$

$$m_2(t) = \tanh\left(\frac{\beta}{2} V_2\right) = \tanh\left(\frac{\beta}{2} \sum_{j=1}^{n} T_{2j} \frac{m_j(t)+1}{2}\right) - \varepsilon \quad (10)$$

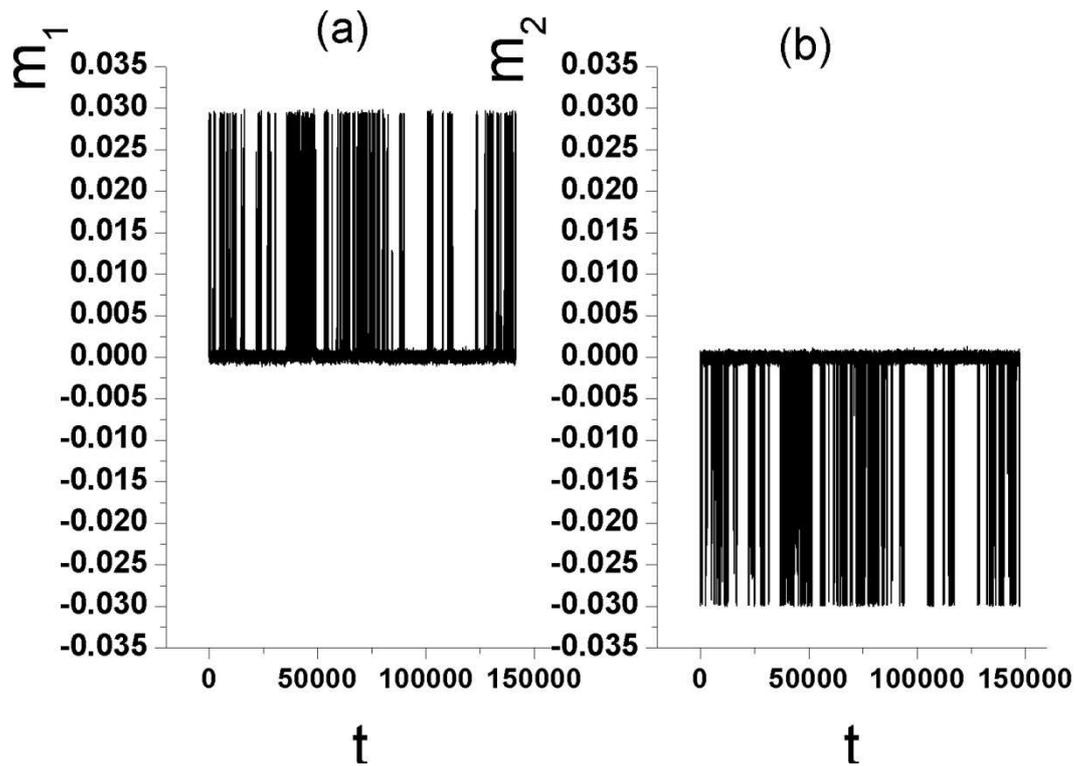

*Figure 2. (a) The time-series of order parameter for excitatory neuron (ANN) (b) The corresponding time-series of inhibitory neuron (ANN).*

In the next Fig.3 we demonstrate the existence of the same structure between the time-series from the ANN model and the corresponding time-series of membrane potential for an excitatory biological neuron, which is in a network connected with other neurons [2].

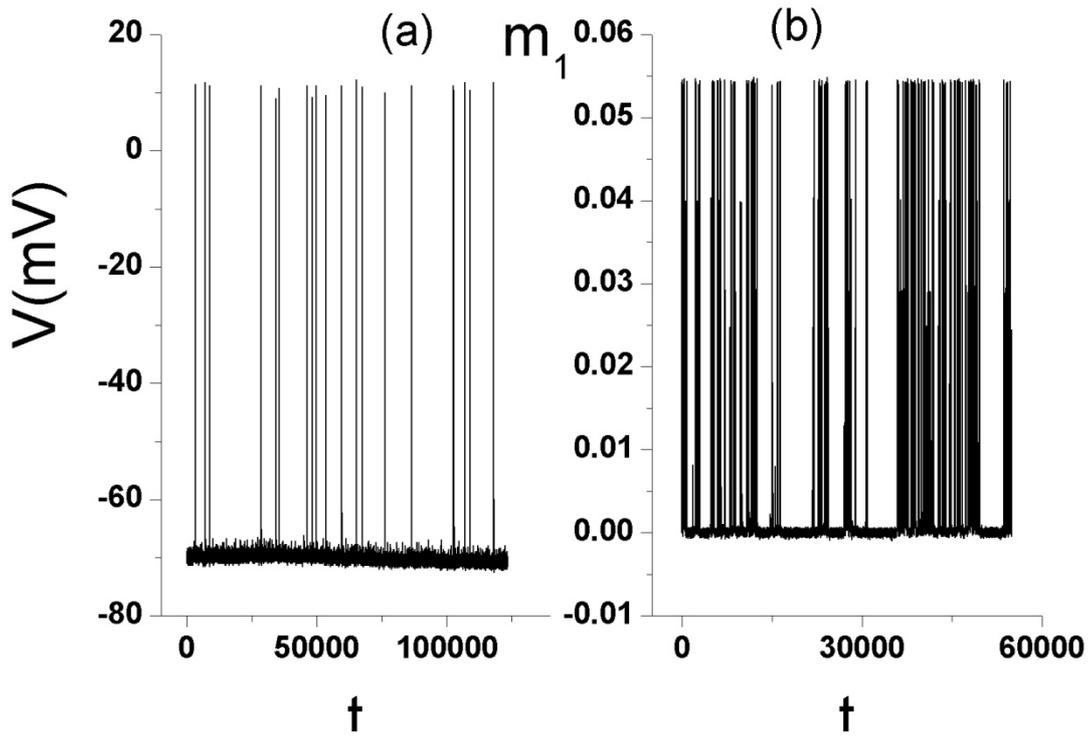

*Figure 3. (a) Amplitude membrane potential fluctuations in vitro intracellular recordings from CA1 pyramidal neurons of Wistar male rats [2]. (b) The time-series of excitatory neuron produced from the proposed ANN model.*

The quantitative description of the dynamics of the fluctuations of the grass is an interesting issue. This is performed using the method of critical fluctuations (MCF). Below we present a short description of this method and refer the reader to the corresponding references for further details.

## 5. The Method of Critical Fluctuations

A way to find the existence of the critical state is the analysis of order parameter fluctuations $\phi$ by the Method of Critical Fluctuations (MCF) . Details for this method are presented in [10]. Importantly, the exact dynamics at the critical point can be determined analytically for a large class of critical systems introducing the so-called critical map . This map can be approximated as an intermittent map:

$$\phi_{n+1} = \phi_n + u\phi_n + \varepsilon_n \qquad (11)$$

The shift parameter $\varepsilon_n$ introduces a non-universal stochastic noise which is necessary for the creation of ergodicity [8]. Each physical system has its characteristic "noise", which is expressed through the shift parameter $\varepsilon_n$. Notice, for thermal systems the exponent z is connected with the isothermal critical exponent $\delta$ as $z=\delta+1$. The crucial observation in this approach is the fact that the distribution $P(l)$ of the suitable defined laminar lengths $l$ (waiting times in laminar region) [7] of the above mentioned intermittent map of Eq. (11) in the limit $\varepsilon_n \to 0$ is given by the power law [11]

$$P(l) \sim l^{-p_l} \qquad (12)$$

where the exponent $p_l$ is connected with the exponent $z$ by $p_l = \frac{z}{z-1}$. Therefore the exponent $p_l$ is connected with the isothermal exponent $\delta$ by: $p_l = 1 + \frac{1}{\delta}$. The distribution of the laminar lengths of fluctuations is fitted by the function:

$$P(l) \sim l^{-p_2} e^{-p_3 l} \qquad (13)$$

We focus on the exponents $p_2$ and $p_3$. If the exponent $p_3$ is zero, then, the exponent $p_2$ is equal to the exponent $p_l$. The relation $p_l = \frac{z}{z-1}$ suggests that the exponent $p_l$ (or $p_2$) should be greater than 1. On the other hand according to the theory of critical phenomena [12] the isothermal exponent $\delta$ is greater than 1. Therefore, as a result from $p_l = 1 + \frac{1}{\delta}$ we take $1 < p_l(p_2) < 2$. In conclusion, the critical profile of the temporal fluctuations is restored by the restrictions: $p_2 > 1$ and $p_3 \approx 0$. As the system removes from the critical state, the exponent $p_2$ decreases while simultaneously the exponent $p_3$ increases reinforcing, in this way, the exponential character of the laminar lengths distribution.

## 6. The results from MCF analysis

The numerical experiment refers to a network of 20 neurons, where we have selected to record the time evolution of the neuron with marker 1, which is in an environment that creates another 19 neurons all of them fully connected. The temperature is T = 2.3. To study the dynamics of fluctuations in the "grass" we cut the spikes thus creating a time series in the interval [=0.0015, 0.0015] because we want to focus on the grass and his does not affect essentially the results. The added noise has width $\varepsilon_o = 0.0004$. The results of the analysis with the MCF are shown in Figure 4.

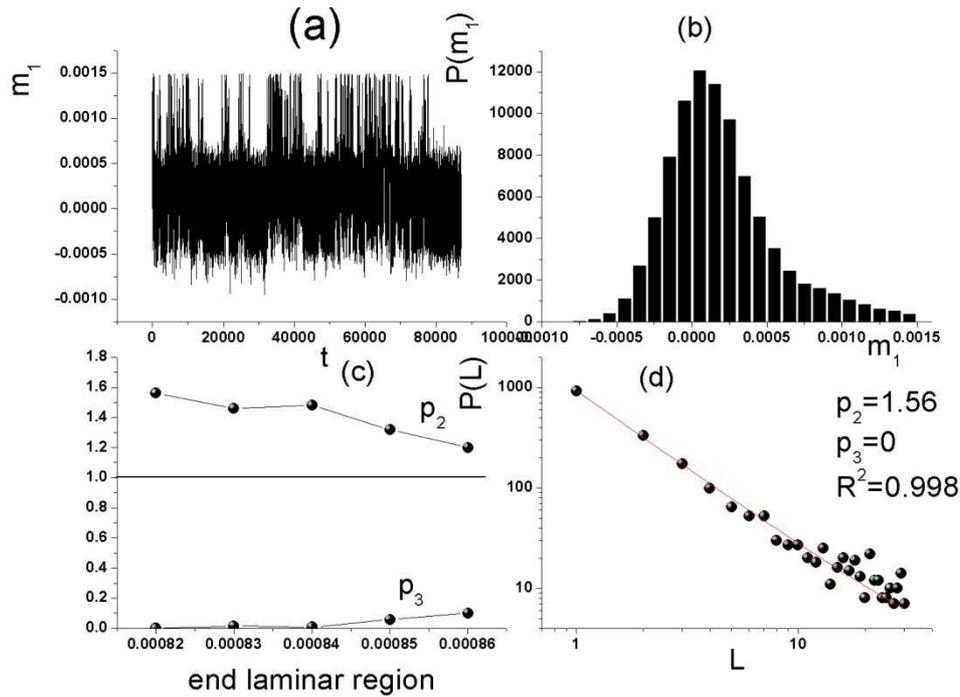

*Figure 4. (a) A $m_1$ timeseries in interval [-0.0015, 0.0015]. (b) The time-series values distribution. (c) The exponents $p_2$, $p_3$ vs the end of laminar region. All these values obey to criticality conditions. (d) A representative laminar distribution in the region [-0.0005, 0.00082] with $p_2 = 1.56, p_3 = 0$, $R^2$=0.998.*

Recently, in our work [2] in order to highlight the dynamics of a neuron found on a network, we analyzed through MCF the recording of the membrane potential (which plays the role of order parameter [2]) of one neuron in an in vitro experiment. The conclusion that we reached was that the neuron when fires, the fluctuations of the membrane potential in relaxation phase between the spikes are critical according to the MCF analysis. This, as we have shown, is a result that is not covered by the H-H theory which very well describes the creation of spikes. Applying this method to biological neurons, an average value to the isothermal critical exponent of δ = 1.89 has been calculated [2], resulting in an average value for exponent $p_2$ = 1.53. As it can be seen from Figure 4, the value of $p_2$ that is close to a power law is 1.56, very close to 1.53 so that the above ANN example is considered to be representative of the dynamics of the neurobiological reality.

## 7. Changes in the parameter of model

There are many factors involved in the issue of dynamics that we have presented. In this work we are investigating some of them. Nevertheless, this issue is open to further study. Before proceeding with such an investigation, however, we will ask an important question. The Metropolis algorithm that we used in our algorithm selects the configurations using the Boltzamn $e^{-\beta H}$ coefficients as statistical weights and not the coefficients of the Fermi distribution $\frac{1}{1+e^{-\beta \varepsilon}}$ due to the order parameter we define. So if we kept the Metropolis algorithm as it is with the Boltzmann statistic weights, then what would that be in our results?

In figure (5) the answer for this question is shown.

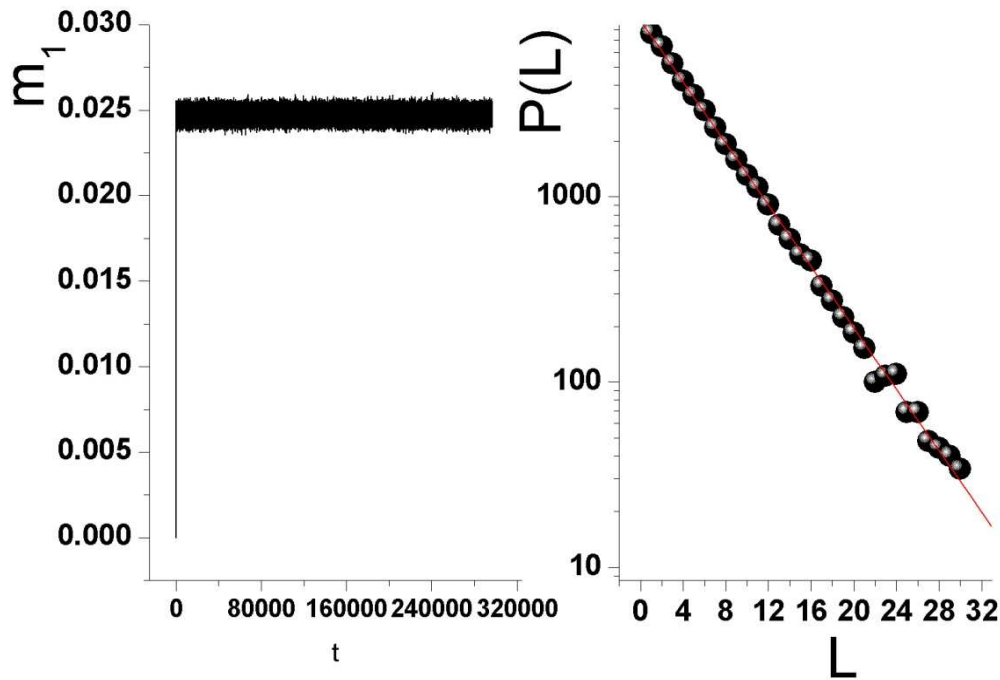

*Figure 5. (a) Changing the statistics from Fermi to Boltzmann the spikes vanished. (b) The distribution of laminar lengths in grass from power-law in Fermi statistics convert to exponential in Boltzmann Statistics.*

The change from Fermi to Boltzmann statistics has eliminated the algorithm's ability to produce the spikes of the dynamic behavior of the neuron as shown in Figure 5a while the distribution of the laminar lengths from a scaling distribution was converted into an exponential distribution according to figure (5b), declaring the absence of any long range dynamics, as occurs in a time series of random numbers. Therefore, the Fermi Statistics is a reason that determines the operation of a neurons network.

The network we tested was 100% complete, all neurons were connected to each other. But a situation closer to reality is when there are no connections between some neurons. Figure 6 shows the results of the MCF analysis when we have a percentage of 20% neurons not connected, so that 20% of the elements in matrix T are zeros in a random way (of course beyond the diagonals is zero). The results are that the critical dynamics of ANN remains as depicted in fig. 6c.

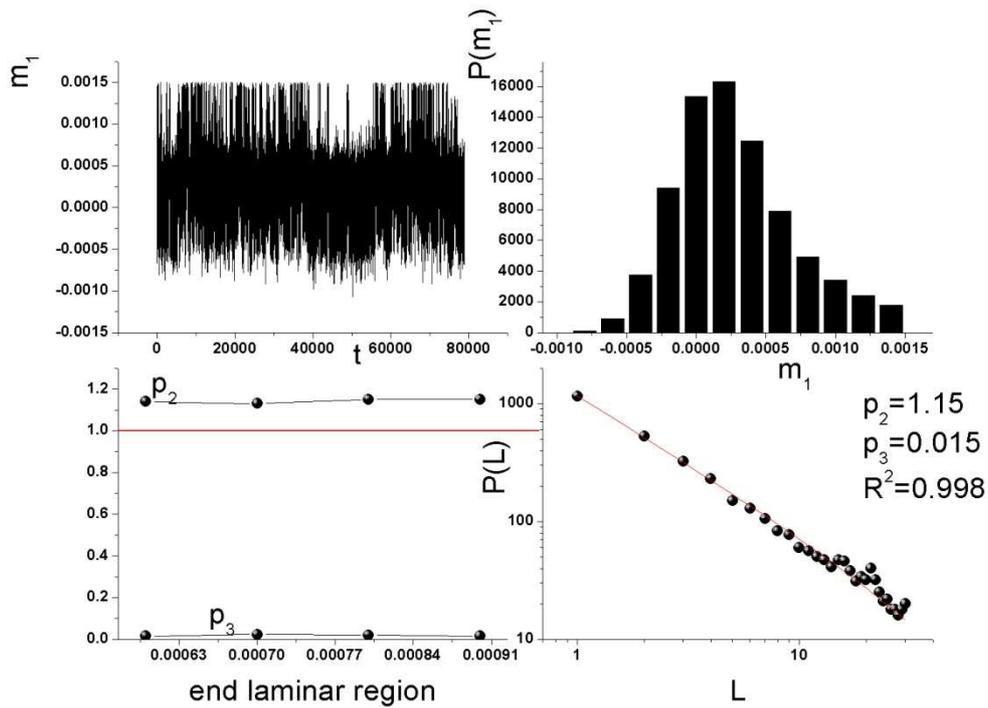

*Figure 6. The critical dynamics remains even if a percentage of neurons synapses are inactive.*

We will now investigate whether the condition that the intensity of inhibitory connections is nearly 4 times stronger than the intensity of the excitatory connections affects the results. In Figure 7 we see this dependence.

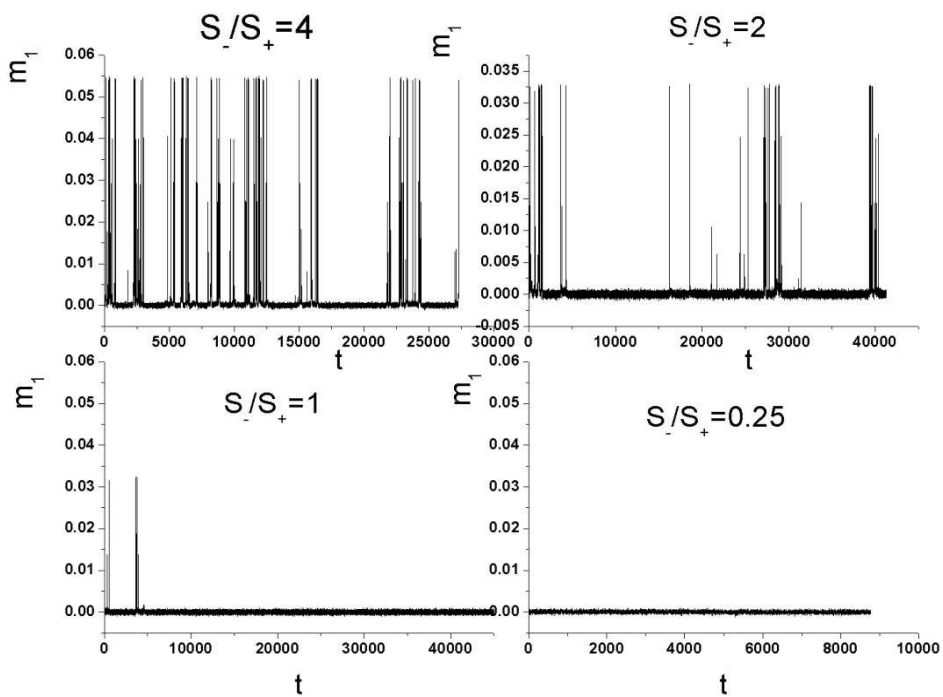

**Figure 7. The neuron timeseries for various values of ratio $\frac{S_-}{S_+}$, which is the ratio of the intensities of inhibitory and excitatory synapses correspondingly.**

We find that the ratio $\frac{S_-}{S_+}$ is decisive for the density of the spikes in the time series. When the ratio decreases, then the density is also reduced until a full inversion of the intensity of the connections takes place. Thus, the spikes disappear, while the analysis with the MCF shows that riticality disappears.

In the analysis we performed we had a number of neurons N = 20. How does the change in the number of neurons affect the results?

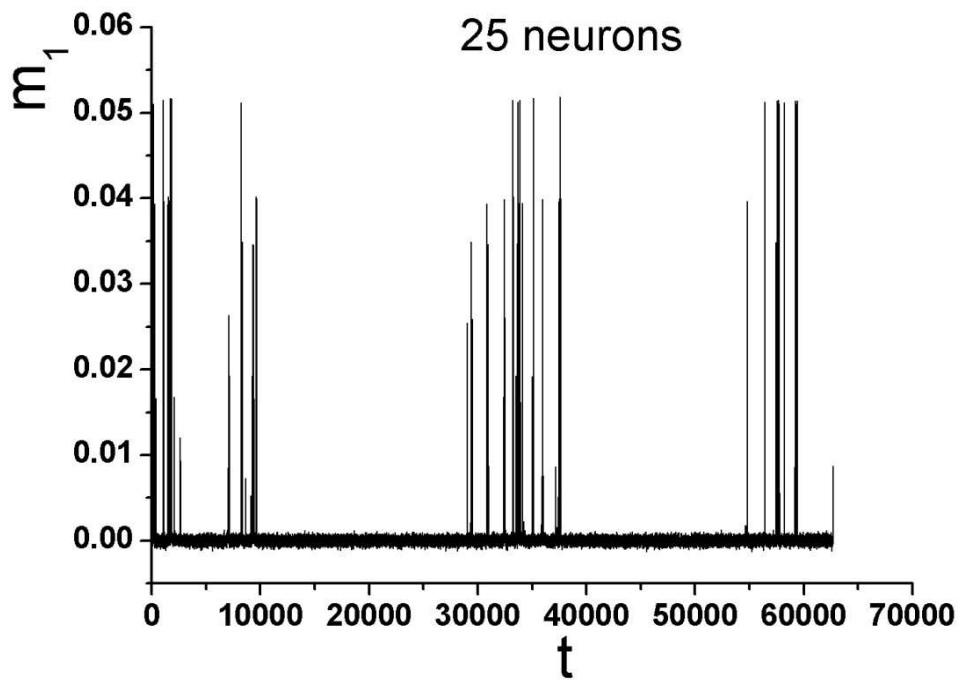

**Figure 8. The time-series of a neuron when the network contains 25 neurons. The density of spikes decreases as the neurons number increases.**

Keeping the temperature and the initial conditions constant we find the time evolution for 25 neurons. In Figure 8 we see that by increasing the number of the neurons, the density of the spikes decreases or otherwise the rate that the neurons fire is reduced. This means that in order to maintain the excitatory function of the neuron, its operating time should be increased.

# 8. Simulation of membrane potential

We are attempting a direct comparison of the biological neuron membrane potential with the corresponding potential of the ANN model $V_1$ which we have shifted by $-0.07\ mV$. The membrane potential given from eq.6 for $i = 1$, where we have imposed the shift $V_1 \to V_1 - 0.07$. The matrix elements $V_{1j}$ is given by eq.5 where we have added the noise $\varepsilon_{1j}$ :

$$V_{1j} = T_{1j} \frac{m_j(t)+1}{2} - \varepsilon_{1j} \quad (14)$$

The eq.7 for the order parameter is not used here.

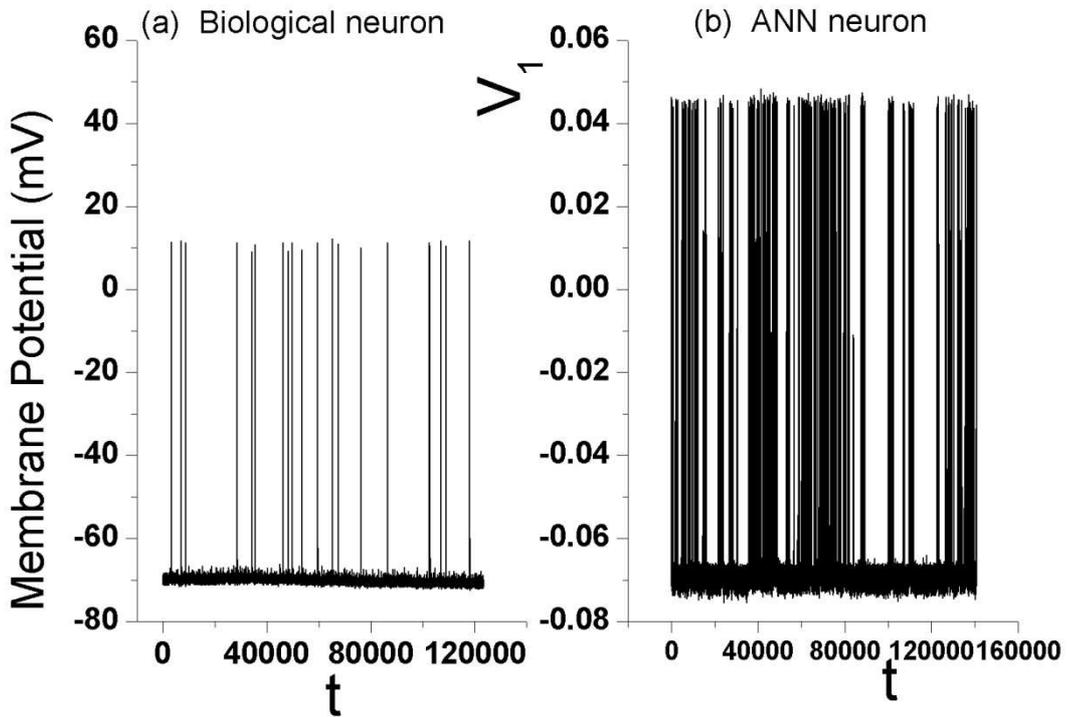

Figure 9. (a) *The timeseries of membrane potential for excitatory biological neuron* (b) The potential $V_1$ of the ANN model. In this case the grass is around -0.07V.

An application of MCF on $V_1$-timeseries gives similar results with the order parameter case. The MCF analysis on $V_1$ time-series shows that the critical behavior remains. In figure 10, details from fig.9 are shown. The structure of the spike as well as the hyper-polarization phase, where the mean level of membrane potential immediately after the spike is lower than the resting potential level are shown clearly. It is characteristic that such a structure like biological action potential appears also in the detail of the ANN simulation (fig.10b).

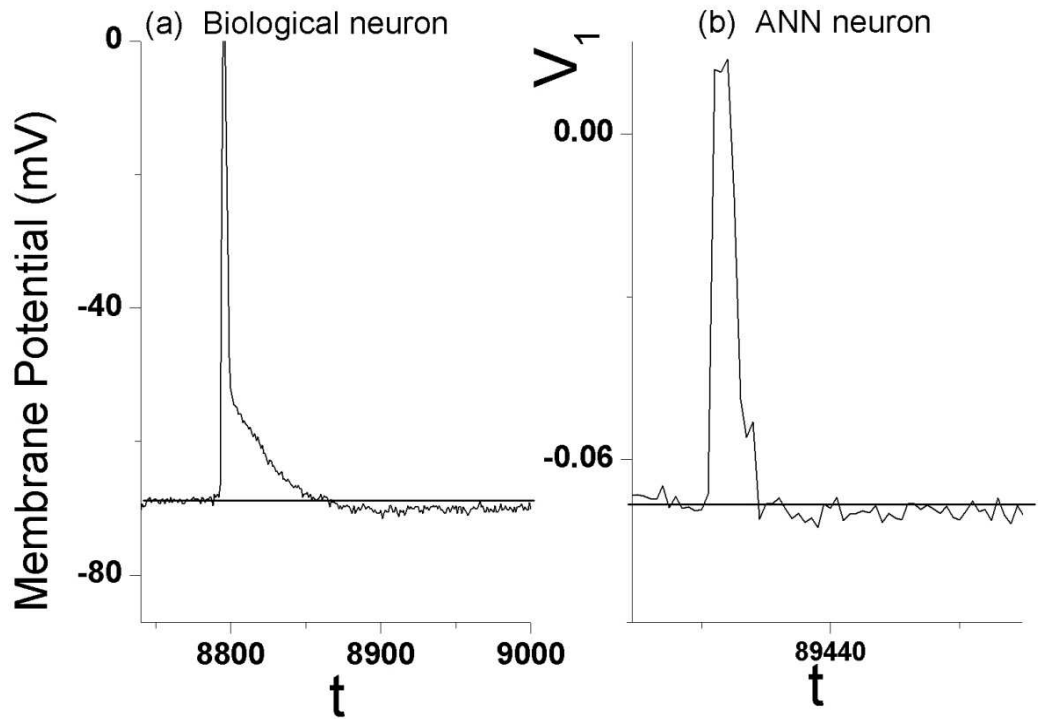

Fig. 10 (a) Detail from membrane potential of neurobiological reality where the spike as well as the hyper-polarization effect are shown. (b) Detail from ANN for the potential $V_1$ as well as the corresponding hyper-polarization effect are also shown.

## 9. Conclusion-Discussion

A simulation for the biological neuron operation is possible through an ANN which operates according to basic principles of Physics, such as the principle of energy minimization as dictated by the second law of thermodynamics. This principle can be considered as an expression of the most important principle of nature (classical or quantum), which is the principle of the least action. This is accomplished numerically through the metropolis Algorithm where the Boltzmann Statistics has been replaced by the Fermi statistics and the interaction of neuron is global with the all connected neurons and not only the nearest neighbors. The verification that this ANN operates as a real neuron is coming when this ANN model accomplishes to produce the same dynamic behavior with the biological neuron in the level of properly defined order parameter as well as the membrane potential. This dynamics is the dynamics which organizes the critical state in the relaxation phase of the biological neuron operation, where the H-H model fails to describe the order parameter fluctuations.

During the biological neuron operation the phase of hyper-polarization appears, where the membrane potential drops under the resting potential. How will the neuron return to its original state of rest? That is, how to restore the initial concentrations of the ions inside and outside the neuron. This job is undertaken by the sodium-potassium ion pump. This pump transfers the ions unlike their spontaneous diffusion. Thus, it transports sodium ions from the inside to the exterior through the sodium channels and restores the initial concentration of sodium ions to the exterior while, through the potassium channels, it transfers potassium ions from the exterior to the inside, restoring the initial potassium concentration to the interior. This phase of restoring the neuron to its resting state occurs within the relaxation time between two spikes is the grass where we find the intermittent dynamics and criticality. Where are these critical fluctuations coming from? Pump operation obeys to dynamic feedback (characteristic of intermittency). Indeed, the rate of pump operation is not constant but is in dialogue with the membrane potential of the neuron (or equivalent to ion concentrations). In other words, the flow of $ions$ through its channel determines the membrane potential and inversely. Therefore, the origin of fluctuations is a feedback process between the pump operation of ions $Na^+, K^+$ channels and membrane potential, which creates the intermittent dynamics. It is important for the consistency of ANN that in this model the hyperpolarization phenomenon appears too (fig.10b).